# Relative roles of multiple scattering and Fresnel diffraction in the imaging of small molecules using electrons, Part II: Differential Holographic Tomography

T.E. Gureyev[*,1,2,3,4], H.M. Quiney[1], A. Kozlov[1], D.M. Paganin[2], G. Schmalz[3] and L.J. Allen[1]

[1)] *ARC Centre in Advanced Molecular Imaging, School of Physics, The University of Melbourne, Parkville 3010, Australia*

[2)] *School of Physics and Astronomy, Monash University, Clayton 3800, Australia*

[3)] *School of Science and Technology, University of New England, Armidale 2351, Australia*

[4)] *Faculty of Health Science, The University of Sydney, Sydney 2006, Australia*

## Abstract

It has been argued that in atomic-resolution transmission electron microscopy (TEM) of sparse weakly scattering structures, such as small biological molecules, multiple electron scattering usually has only a small effect, while the in-molecule Fresnel diffraction can be significant due to the intrinsically shallow depth of focus. These facts suggest that the three-dimensional reconstruction of such structures from defocus image series collected at multiple rotational orientations of a molecule can be effectively performed for each atom separately, using the incoherent first Born approximation. The corresponding reconstruction method, termed here Differential Holographic Tomography, is developed theoretically and demonstrated computationally on several numerical models of biological molecules. It is shown that the method is capable of accurate reconstruction of the locations of atoms in a molecule from TEM data collected at a small number of random orientations of the molecule, with one or more defocus images per orientation. Possible applications to cryogenic electron microscopy and other areas are briefly discussed.

## 1. Introduction

Cryogenic electron microscopy (cryo-EM) has recently reached the true atomic spatial resolution milestone [1-2]. As the spatial resolution in high-energy transmission EM (TEM) approaches one ångström, the corresponding depth of focus (also known as the depth of field) tends to become shallower than the longitudinal (along the direction of propagation of the illuminating electron wave) extent of a typical imaged sample, such as a protein molecule. For example, at a spatial resolution of $\Delta = 1$ Å and a wavelength of $\lambda = 0.025$ Å (for electrons at 200 keV energy), the depth of focus is equal to $z = \Delta^2 / (2\lambda) = 20$ Å [3]. It has been suggested previously that, as a result of the shallow focus, the in-molecule Fresnel diffraction (free-space propagation) becomes a significant factor that cannot be ignored in the reconstruction of the sample from its defocused images [3-4]. In other words, the propagation of the incident electron wave through the sample can no longer be satisfactorily approximated by projection integrals of the electrostatic potential along straight lines. Instead, the Fresnel diffraction of the electron beam inside the molecule must be explicitly taken into account in order to achieve an accurate 3D reconstruction. The introduction of specially designed aberrations into the EM illumination system can increase the depth of focus [5-6], but this is still unlikely to fully eliminate the in-molecule Fresnel diffraction effects at atomic resolution. Moreover, as we argue below (see also [7]), attempts to increase the depth of focus in EM tomography may in some cases be counter-productive, since a shallow focus improves the spatial resolution by enhancing the longitudinal localization of individual atoms in a molecule.

[*] Corresponding author. *E-mail address*: timur.gureyev@unimelb.edu.au (T.E. Gureyev).

It is known that, in the transmission electron microscopy of sufficiently thin non-crystalline specimens, "diffraction tomography" (DT) [8-10] can be used to properly account for the Fresnel diffraction inside the sample and correctly reconstruct the three-dimensional (3D) distribution of the complex refractive index, or, equivalently, the spatial distribution of the electrostatic potential in electron imaging. The DT approach can be based on the first Rytov or first Born approximation, instead of the projection approximation utilized in conventional computed tomography (CT). The use of DT for 3D reconstruction of small molecules was discussed in detail in the first part of this work [7]. Here, we develop and test a different method, termed "Differential Holographic Tomography" (DHT), aimed at reconstruction of the 3D structure of small molecules from TEM defocus series collected at a small number of different orientations of the molecule. This method is developed in detail in the next section, and here we only briefly discuss the main physical considerations that underpin this method.

We showed in [7] that, in agreement with the statements found in [3-4] and elsewhere, in high-resolution electron microscopy (EM) of "sparsely localized" structures, such as small biological molecules, the total contrast distribution in a defocused image can typically be well approximated by an incoherent sum of contrast functions corresponding to individual atoms. This fact can be used as a basis for reconstructing the 3D distribution of the electrostatic potential in a molecule from transmission images collected at different defocus distances and different orientations of the molecule, effectively performing the reconstruction separately for each atom before adding together the individual reconstructed atomic potentials. This general approach is similar to the "independent atoms model" which has been used for crystallographic phase retrieval, sometimes in conjunction with the atomicity constraint, for a long time (see e.g. [11-12]). However, we note that the independent atoms approximation, which ignores molecular effects, is often applied in a far-field diffraction regime, where the interference of the waves scattered by different atoms is essential. In contrast, the present method is concerned with the Fresnel diffraction region and ignores the interference effects, thus using an incoherent independent atoms approximation. The validity of such an approximation was studied previously in the context of the first Born approximation [7] and is also considered in the next section of this paper.

In order to be able to perform independent localized reconstruction of the potential in the vicinity of each atom, one can collect, for a given orientation of the molecule, one or more defocused images along the propagation direction of the transmitted electron wave. These images can be used to perform computational phase retrieval, e.g. with the help of the Iterative Wave Function Reconstruction (IWFR) method from [13] or the single-image phase retrieval method from [14]. The complex wave amplitude thus reconstructed at a given rotational orientation of the molecule can then be numerically back-propagated to the vicinity of each atom and the contrast function can be evaluated in that vicinity. As the local distribution of the electrostatic potential in the neighborhood of an atomic nucleus is essentially spherically symmetric, its reconstruction from the localized contrast function becomes relatively trivial. In particular, we will show that such a reconstruction can be obtained by summing up slightly defocused local contrast functions obtained at different orientations (that is, averaging them over the illumination directions) and then taking the inverse 3D Laplacian of the sum. The latter operation effectively compensates for the local Transport of Intensity equation (TIE) type differential phase contrast and recovers the potential from its Laplacian [15]. The differential nature of this method for reconstruction of the electrostatic potential is essential in the case of phase objects, such as atoms illuminated by high-energy electron waves, since they do not produce intensity contrast in the exact in-focus position.

By construction, the electrostatic potential reconstructed by the proposed method is localized to narrow regions around the atomic positions along the direction of propagation of the electron wave through the molecule at each orientation. This localization property of the reconstruction algorithm allows one to significantly relax the sampling conditions with respect to the number of required rotational orientations. In the case of conventional CT [16], the usual Nyquist conditions relate the required number of angular steps, $n_a$, over 180 degrees of rotation to the number of pixels in the detector rows, $n_x$, via the well-known sampling relationship $n_a = (\pi/2)n_x$. In contrast, in our DHT algorithm, the number of required angular positions becomes almost independent of the number of detector pixels, due to the longitudinal localization of the back-propagated contrast function. Indeed, the relevant number of rotational positions here is related to the size of the vicinity of an individual atom inside which its electrostatic potential is significant, instead of being related to the size of the whole molecule. The quality of reconstruction using the DHT method is still likely to depend on the size of the molecule. However, the primary mechanism for the deterioration of the DHT reconstruction quality with the increase in the size of the molecule is in this case related primarily to the increase of the contribution of multiple scattering effects to the contrast function. These effects include the "shading" of atoms by one another along the lines of tomographic projections. The negative influence of these multiple scattering effects on the reconstruction quality can be alleviated by increasing the number of angular orientations in the scan. Several papers have been published recently on the incorporation of multiple scattering into methods for 3D EM reconstruction [17-19]. Potentially, some of these techniques could be used in conjunction with DHT in the future to account for the effects of multiple scattering in the input data and make the resultant algorithm more robust in this respect.

Apart from its connections with DT [8-10] stated above, the proposed DHT method has some similarities with the Big Bang Tomography technique developed by D. Van Dyck and co-authors [20-22]. In that technique, the "depth" position (i.e. the position along the direction of the optical axis) of different atoms in the imaged sample is determined using the relevant contrast transfer function (CTF), which is similar in principle to the approach used in DHT. However, the reconstruction technique used in [20-22] is different from that of DHT. Another method, known as holotomography [23-24], also utilizes multiple defocused images to retrieve the complex wave amplitude in the Fresnel region before the subsequent numerical back-propagation and tomographic reconstruction. Unlike the DHT method developed below, however, holotomography employs conventional CT for reconstruction of the 3D distribution of the real and imaginary parts of the refractive index in the imaged sample from the recovered defocused complex amplitudes.

## 2. Differential holographic tomography algorithm

Consider an imaging setup with a monochromatic plane wave $I_{in}^{1/2}\exp(i2\pi kz)$ illuminating a weakly scattering object, where $k=1/\lambda$ is the wave number, $I_{in}=\text{constant}$ is the intensity of the incident wave and $\mathbf{r}\equiv(x,y,z)$ is a Cartesian coordinate system in 3D space. The complex amplitude $U(\mathbf{r})$ of the wave inside the object satisfies the time-independent Schrödinger equation: $\nabla^2 U(\mathbf{r})+4\pi^2 n^2(\mathbf{r})k^2 U(\mathbf{r})=0$, where $n(\mathbf{r})$ is the refractive index. In the case of electron microscopy, one has $n(\mathbf{r})\cong 1+V(\mathbf{r})/(2E)$, where $V(\mathbf{r})\geq 0$ is the electrostatic potential, $E$ is the accelerating voltage and $\varepsilon\equiv V(\mathbf{r})/(2E)$ is typically much less than unity [25]. We consider the problem of reconstruction of the 3D distribution of the electrostatic potential from the intensity of transmitted waves measured at some distances from the object (defocused images), for a number of different orientations of the object.

For configurations in which the object is thicker than the depth of focus, but is so weakly scattering that the multiple scattering can be safely ignored, the first Born approximation can be applied. In this case, the change of the incident wave due to propagation through the object can be neglected and the intensity of the projection image collected at a position, $z$, downstream from the object along the optical axis can be expressed as an incoherent sum of the primary beam intensity and the intensities scattered by each atom independently when illuminated by the unperturbed incident plane wave [7,26]. The terms corresponding to the interference of the waves scattered by different atoms are of the second order ($\varepsilon^2$) with respect to the implicitly small parameter, $\varepsilon=V(\mathbf{r})/(2E)$, in the Born series and, hence, can be neglected. In particular, it has been shown in [7] that, using the first Born approximation for the diffracted wave, the 2D Fourier transform of the intensity of a defocused image can be written as

$$(\mathbf{F}_2 I)(\mathbf{q}_\perp, z) / I_{in} = \delta(\mathbf{q}_\perp) + [2\pi/(\lambda E)] \int \sin[\pi\lambda(z-z')q_\perp^2](\mathbf{F}_2 V)(\mathbf{q}_\perp, z')dz', \qquad (1)$$

where $(\mathbf{F}_2 I)(\mathbf{q}_\perp, z)$ is the 2D Fourier transform of the intensity $I(\mathbf{r}_\perp, z) \equiv |U(\mathbf{r}_\perp, z)|^2$ of the transmitted electron wave, $\mathbf{r} \equiv (\mathbf{r}_\perp, z)$, $\mathbf{r}_\perp \equiv (x, y)$, $\mathbf{q} \equiv (\mathbf{q}_\perp, q_z)$, $\mathbf{q}_\perp \equiv (q_x, q_y)$ and $q_\perp \equiv |\mathbf{q}_\perp|$. The 2D Fourier transform is defined by $(\mathbf{F}_2 f)(\mathbf{q}_\perp) \equiv \iint \exp[-i2\pi\mathbf{q}_\perp \mathbf{r}_\perp] f(\mathbf{r}_\perp)d\mathbf{r}_\perp$. Equation (1) has the form of an incoherent sum of the well-known expressions for the first Born approximation to the scattered intensities [27] corresponding to different transverse planes, $z'$, inside the imaged object.

The electrostatic potential in a molecule can be approximated by a sum of electrostatic potentials associated with the constituent atoms, $V(\mathbf{r}) \cong \sum_{m=1}^{M} V_m(\mathbf{r} - \mathbf{r}^{(m)})$, where $m$ enumerates the atoms in the molecule and $\mathbf{r}^{(m)} \equiv (\mathbf{r}_\perp^{(m)}, z^{(m)})$ are the locations of individual atoms. According to the "independent atoms" model discussed above, the contrast function, $K(\mathbf{r}_\perp, z) \equiv 1 - I(\mathbf{r}_\perp, z)/I_{in}$, produced by the molecule can be approximated by a sum of contrast functions produced by individual potentials $V_m(\mathbf{r} - \mathbf{r}^{(m)})$, resulting in $K(\mathbf{r}) \cong \sum_{m=1}^{M} K_m(\mathbf{r} - \mathbf{r}^{(m)})$. Applying Eq. (1) to each atom separately, we obtain

$$(\mathbf{F}_2 K_m)(\mathbf{q}_\perp, z) = [2\pi/(\lambda E)] \int \sin[\pi\lambda(z'-z)q_\perp^2](\mathbf{F}_2 V_m)(\mathbf{q}_\perp, z')dz'. \qquad (2)$$

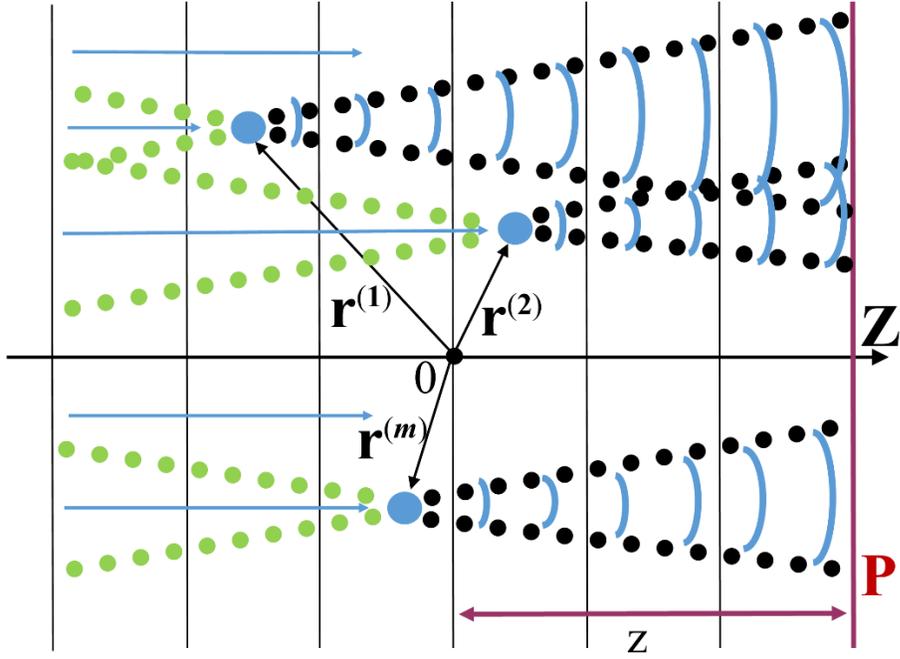

Fig. 1. Schematic representation of the imaging setup.

The imaging setup corresponding to Eqs. (1) and (2) is represented schematically by Fig. 1. Incident monochromatic plane waves scatter from a molecule composed of $M$ atoms centered at position vectors $\mathbf{r}^{(m)} \equiv (\mathbf{r}_\perp^{(m)}, z^{(m)})$, $m = 1, 2,..., M$, relative to an origin $O$. Since the first Born approximation has been assumed, there is a strong unscattered plane wave downstream of the molecule, as indicated by the parallel black vertical lines that fill the diagram. The scattered waves, originating from each atom, have been indicated with curved blue wave-fronts that are bounded by the black dotted lines in the paraxial approximation. Since in-molecule diffraction cannot be neglected, the scattered wave-fronts spread a non-negligible transverse extent, within the volume occupied by the molecule. For example, the wave scattered from the atom at $\mathbf{r}^{(1)}$ has spread a non-negligible transverse amount, once it has propagated to the plane of constant $z$ that contains the centre of the atom located at $\mathbf{r}^{(2)}$. The intensity of the complex wave-field downstream of the molecule is measured over the plane **P**, together with one or more additional parallel planes downstream of **P**. This through-focal-series of intensity maps can then be used to determine the complex wave-field over the plane **P**. Having recovered the complex wave-field over **P**, it can be back-propagated numerically. This generates a coherent superposition of the following three complex wave-fields throughout a compact volume containing the molecule: (i) the unscattered "background" plane wave, indicated by the parallel vertical lines in the diagram of Fig. 1; (ii) the advanced complex wave-field scattered from the molecule, which diverges as it propagates downstream of each atom, as indicated by the dotted black lines; (iii) the retarded complex scattered wave-field, which converges as it propagates upstream of each atom and thereby converges upon the centre of each atom, as indicated by the dotted green lines. Since the scattering potentials for each atom are real, which implies each atom to be a "pure phase" object, and the first Born approximation is applicable, we conclude that the contrast in the numerically back-propagated field is necessarily weak around the atomic positions. This issue is treated via the "differential holographic contrast" device introduced below.

As was shown, for example, in [7], the contrast function obtained with the opposite direction of propagation of the incident plane electron wave is equal to

$$(\mathbf{F}_2 K_{m,\pi})(\mathbf{q}_\perp^-, z) = [2\pi/(\lambda E)] \int \sin[\pi\lambda(-z'-z)q_\perp^2](\mathbf{F}_2 V_m)(\mathbf{q}_\perp, z') dz', \tag{3}$$

where $\mathbf{q}_\perp^- \equiv (-q_x, q_y)$ is the mirror-reflection of the vector $\mathbf{q}_\perp = (q_x, q_y)$ with respect to the rotation axis. For small values of the argument we can approximate the sine functions in the usual way:
$\sin[\pi\lambda(z'-z-d)q_\perp^2] + \sin[\pi\lambda(-z'+z-d)q_\perp^2] \cong \pi\lambda(z'-z-d)q_\perp^2 + \pi\lambda(-z'+z-d)q_\perp^2 = -2\pi\lambda d q_\perp^2$,
where $d$ is a specially introduced "local defocus" parameter, the role of which will become clear below. Let us introduce the local "symmetrized" defocus contrast function
$K_{d,m}(\mathbf{r}_\perp, z) \equiv K_m(\mathbf{r}_\perp, z+d) + K_{m,\pi}(\mathbf{r}_\perp^-, -z+d)$, with $\mathbf{r}_\perp^- \equiv (-x, y)$. Note that the transverse localization of the function $K_{d,m}(\mathbf{r}_\perp, z)$ is usually much narrower than the longitudinal localization, as can be seen, for example, in the case of explicit expressions for a Gaussian model of the electrostatic potential given in the Appendix. The essence of the "independent atoms" model is in the assumption that the supports (areas with non-zero values) of the functions $K_{d,m}(\mathbf{r}_\perp - \mathbf{r}_\perp^{(m)}, z - z^{(m)})$ with different indices $m$ (that is, corresponding to different atoms) overlap only over the areas where the potential values are negligibly small, regardless of the illumination direction.

Since the potential $V_m(\mathbf{r})$ is highly localized around $\mathbf{r} = 0$, Eqs. (2) and (3) allow the following approximation at small values of the defocus parameter $d$:

$$(\mathbf{F}_2 K_{d,m})(\mathbf{q}_\perp, 0) \cong -(d/E) 4\pi^2 q_\perp^2 \int (\mathbf{F}_2 V_m)(\mathbf{q}_\perp, z') dz' = -(d/E) 4\pi^2 (q^2 \mathbf{F}_3 V_m)(\mathbf{q}_\perp, 0). \tag{4}$$

This relationship remains true regardless of the direction of the one-dimensional integral (line projection) in 3D, meaning that the two-dimensional function $K_{d,m}(\mathbf{r}_\perp, 0)$ is always equal to the one-dimensional projection of the three-dimensional function $(d/E)\nabla^2 V_m(\mathbf{r})$, provided that $\mathbf{r}_\perp$ represents the two-dimensional Cartesian coordinates in the plane orthogonal to the direction of the projection:

$$K_{d,m}(\mathbf{r}_\perp, 0) \cong (d/E)\nabla_\perp^2 (\mathbf{P}_1 V_m)(\mathbf{r}_\perp) = (d/E)(\mathbf{P}_1 \nabla^2 V_m)(\mathbf{r}_\perp), \quad (5)$$

where $\mathbf{P}_1$ denotes the one-dimensional projection in the direction orthogonal to $\mathbf{r}_\perp$. Equation (4) has the form of the classical projection slice theorem [16]. Therefore, if the function $K_{d,m}(\mathbf{r}_\perp, 0)$ can be obtained from experimental measurements at different orientations (rotational positions) of the molecule, this data can be used to reconstruct the local 3D distribution $(d/E)\nabla^2 V_m(\mathbf{r})$ and, consequently, the local 3D potential $V_m(\mathbf{r})$.

In order to find the function $K_{d,m}(\mathbf{r}_\perp, 0)$ in a vicinity of an atom location $\mathbf{r}^{(m)} \equiv (\mathbf{r}_\perp^{(m)}, z^{(m)})$, from experimental measurements, one can measure several defocus images along the direction $z$, perform phase retrieval (e.g. using the IWFR method from [13] or the single-image phase retrieval method from [14]) and then numerically back-propagate the obtained complex amplitude to the planes $z = z^{(m)} \pm d$, as needed for evaluation of the function $K_{d,m}(\mathbf{r}_\perp - \mathbf{r}_\perp^{(m)}, z - z^{(m)})$.

Applying the usual filtered back-projection (FBP) method [16] to Eq. (5), allows one to reconstruct the 3D distribution of $V_m(\mathbf{r})$ from the left-hand side $K_{d,m,\theta}(\mathbf{r}_\perp, 0)$ obtained at different angular positions $\theta$ with respect to rotations of the atom around the $y$ axis:

$$V_m(\mathbf{r}) = \frac{E}{d}\nabla^{-2}\int_0^\pi \int_{-\infty}^\infty \int_{-\infty}^\infty \exp(-i2\pi \mathbf{r}_{\theta,\perp} \mathbf{q}_\perp)(\mathbf{F}_2 K_{d,m,\theta})(\mathbf{q}_\perp, 0)|q_x|\, d\mathbf{q}_\perp d\theta, \quad (6)$$

where $(\mathbf{F}_2 K_{d,m,\theta})(\mathbf{q}_\perp, 0)$ is the two-dimensional Fourier transform of the function $K_{d,m,\theta}(\mathbf{r}_{\theta,\perp}, 0)$ and $\theta$ is the angle between the original $z$ axis and the rotated $x_\theta$ axis, $x_\theta = x\sin\theta + z\cos\theta$, $\mathbf{r}_{\theta,\perp} \equiv (x_\theta, y)$ and $z_\theta = -x\cos\theta + z\sin\theta$. Note that the angle between the original $z$ or $x$ axes and the corresponding rotated $z_\theta$ and $x_\theta$ axes, respectively, is equal to $\theta' = \pi/2 - \theta$.

We now turn to the reconstruction of the electrostatic potential for the whole molecule. Since

$$V(\mathbf{r}) = \sum_{m=1}^{M} V_m(\mathbf{r} - \mathbf{r}^{(m)}) \text{ and}$$

$$\sum_{m=1}^{M} K_{d,m}(\mathbf{r}_\perp - \mathbf{r}_\perp^{(m)}, 0) \cong \sum_{m=1}^{M} K_{d,m}(\mathbf{r}_\perp - \mathbf{r}_\perp^{(m)}, z - z^{(m)}) = \sum_{m=1}^{M} K_m(\mathbf{r}_\perp - \mathbf{r}_\perp^{(m)}, z - z^{(m)} + d)$$

$$+ K_{m,\pi}(\mathbf{r}_\perp^- - \mathbf{r}_\perp^{(m)}, -z + z^{(m)} + d) \cong K(\mathbf{r}_\perp, z + d) + K_\pi(\mathbf{r}_\perp^-, -z + d),$$

Eq. (6) leads to the following reconstruction formula for the 3D distribution of the electrostatic potential in the whole molecule:

$$V(\mathbf{r}) = \frac{E}{d} \nabla^{-2} \sum_{m=1}^{M} \int_0^\pi \int_{-\infty}^{\infty} \int_{-\infty}^{\infty} \exp[-i2\pi(\mathbf{r}_{\theta,\perp} - \mathbf{r}_\perp^{(m)}) \mathbf{q}_\perp] (\mathbf{F}_2 K_{d,m,\theta})(\mathbf{q}_\perp, 0) |q_x| d\mathbf{q}_\perp d\theta$$

$$\cong \frac{E}{d} \nabla^{-2} \int_{-\pi}^{\pi} \int_{-\infty}^{\infty} \int_{-\infty}^{\infty} \exp(-i2\pi \mathbf{r}_{\theta,\perp} \mathbf{q}_\perp) \mathbf{F}_2 K_\theta(\mathbf{q}_\perp, z_\theta + d) |q_x| d\mathbf{q}_\perp d\theta. \tag{7}$$

In the derivation of Eq. (7) we have used the same (optional) symmetry considerations as in our derivation of Eq. (10) in [7], which allowed us to replace the function $\mathbf{F}_2 K_{\pi+\theta}(q_x, q_y, z_{\pi+\theta} + d)$ with $\mathbf{F}_2 K_\theta(-q_x, q_y, -z_\theta + d)$ by extending the angular integration over $\theta$ to the additional semi-circle $(-\pi, 0]$. Note that while Eq. (7) superficially resembles Eq. (10) from [7], it is in fact quite different, due to the presence of the third argument, $z_\theta + d$, in the contrast function under the integrals. This feature points to the highly localized nature of this reconstruction algorithm: the contribution to the local value of the reconstructed potential $V(x, y, z)$ comes predominantly from the contrast functions $K_\theta(x \sin\theta + z \cos\theta, y, d - x \cos\theta + z \sin\theta)$ evaluated in the vicinity of the same point $(x, y, z)$. In the FBP algorithm of conventional CT [16], the measured 2D contrast functions, $K_{CT}(\mathbf{r}_\perp) \equiv -\ln[I_1(\mathbf{r}_\perp)/I_0] \cong 1 - I_1(\mathbf{r}_\perp)/I_0$, are effectively spread back uniformly along the rays passing through the sample and ending at the corresponding points $\mathbf{r}_\perp$ at each rotational position. This operation is highly non-local. In contrast, the uniform spreading is replaced in Eq. (7) by the Fresnel back-propagation of the reconstructed complex amplitude, evaluation of the local defocused contrast function $K_\theta(x_\theta, y_\theta, z_\theta + d)$ and the local reconstruction corresponding to Eq. (6). This reconstruction is effectively performed for each atom independently.

An additional important relevant consideration is that the functions $V_m(\mathbf{r})$ can be considered spherically-symmetric, that is $V_m(\mathbf{r}) = V_m(|\mathbf{r}|)$, and hence their projections $(\mathbf{P}_1 V_m)(\mathbf{r}_\perp) \equiv \int V_m(\mathbf{r}_\perp, z')dz'$ are circularly-symmetric: $(\mathbf{P}_1 V_m)(\mathbf{r}_\perp) = (\mathbf{P}_1 V_m)(|\mathbf{r}_\perp|)$. In this case it is known that the 3D potential can be reconstructed from a single projection [16]. As the 3D Laplacian is also spherically symmetric, the previous statement applies to the function $\nabla^2 V_m(\mathbf{r})$ as well. This is particularly easy to see in the Fourier representation, where $(\mathbf{F}_3 \nabla^2 V_m)(\mathbf{q}) = -4\pi^2 |\mathbf{q}|^2 (\mathbf{F}_3 V_m)(\mathbf{q})$. The spherical symmetry of the last function allows us to write $|\mathbf{q}|^2 (\mathbf{F}_3 V_m)(\mathbf{q}) = |\mathbf{q}|^2 (\mathbf{F}_3 V_m)(|\mathbf{q}|, 0, 0)$. The Fourier slice theorem (for the case of rotation around the $y$ axis) then implies $(\mathbf{F}_3 V_m)(|\mathbf{q}|, 0, 0) = (\mathbf{F}_2 \mathbf{P}_1 V_m)(|\mathbf{q}|, 0)$. Finally, we also have that $-4\pi^2 |\mathbf{q}|^2 (\mathbf{F}_2 \mathbf{P}_1 V_m)(|\mathbf{q}|, 0) = (\mathbf{F}_2 \nabla_\perp^2 \mathbf{P}_1 V_m)(|\mathbf{q}|, 0)$. All of this means that, in principle, the 3D distribution of the local potential can be obtained from a single projection, collected at an arbitrary orientation, according to the equation $(\mathbf{F}_3 \nabla^2 V_m)(\mathbf{q}) = (\mathbf{F}_2 \nabla_\perp^2 \mathbf{P}_1 V_m)(|\mathbf{q}|, 0)$. Using Eq. (5), we now obtain:

$$V_m(\mathbf{r}) = (E/d) \nabla^{-2} \mathbf{F}_3^{-1}[(\mathbf{F}_2 K_{d,m})(|\mathbf{q}|, 0, 0)](\mathbf{r}), \qquad (8)$$

where the function $K_{d,m}(\mathbf{r}_\perp, 0) \equiv K_{d,m,\theta}(\mathbf{r}_{\theta,\perp}, 0)$ is the same for any value of $\theta$.

Alternatively, applying the well-known Abel transform formula for the inversion of the 2D Radon transform of circularly symmetric (radial 2D functions) from [16] to the function $K_{d,m}(x, y, 0)$ with respect to the $x$ coordinate at each fixed value of $y$ (i.e. the Abel transform is applied to projections in the $x$-$z$ plane), produces a local reconstruction formula which has a simpler form compared to the generic FBP Eq. (6):

$$V_m\left(\sqrt{\rho^2 + y^2}\right) = \frac{-E}{\pi d} \frac{1}{\rho} \frac{\partial}{\partial \rho} \int_\rho^\infty \frac{\rho' \nabla_\perp^{-2} K_{d,m}(\rho', y, 0)}{(\rho'^2 - \rho^2)^{1/2}} d\rho', \quad \rho = \sqrt{x^2 + z^2}. \qquad (9)$$

As an example, consider an atomic potential approximated by a symmetric Gaussian distribution, $V_m(\mathbf{r}) = 2Ec_m (2\pi)^{-3/2} \sigma_m^{-3} \exp[-|\mathbf{r}|^2 /(2\sigma_m^2)]$, where $c_m$ and $\sigma_m$ are suitable constants with dimensionality of volume and length, respectively. For this potential, it is possible to analytically calculate the corresponding local contrast function along any direction (see Appendix), with the result being $K_{d,m}(\mathbf{r}_\perp, 0) = dc_m / (\pi \sigma_m^2) \nabla_\perp^2 \exp[-|\mathbf{r}_\perp|^2 /(2\sigma_m^2)]$. In this case, the integral over $\rho'$ in Eq. (9) can be evaluated explicitly, resulting in

$$V_m\left(\sqrt{x^2 + y^2 + z^2}\right) \cong \frac{2E}{(2\pi)^{3/2}} \frac{c_m}{\sigma_m^3} \exp\left[\frac{-(x^2 + y^2 + z^2)}{2\sigma_m^2}\right], \tag{10}$$

which obviously represents an exact reconstruction of the original 3D distribution of the atomic potential. It is also straightforward to verify that Eq. (8) exactly reconstructs the Gaussian potential $V_m(\mathbf{r})$ from its contrast function $K_{d,m}(\mathbf{r}_\perp, 0)$.

Since Eq. (9) still has a somewhat complicated form and contains a singular integral, it is useful to simplify it further. We shall do so now with the help of an apparently rather crude approximation which, however, is exact in the case of a Gaussian local potential $V_m(\mathbf{r})$ and, therefore, can be expected to work reasonably well for realistic atomic potentials. Let us assume that the function $(\nabla_\perp^{-2} K_{d,m})(\rho', y, 0)$ is zero when $\rho' > \rho_m$ and is sufficiently slowly varying with respect to the argument $\rho$ when $\rho < \rho' < \rho_m$, so that it can be taken outside of the integral in Eq. (9) and the subsequent derivative with respect to $\rho$. The parameter $\rho_m$ is associated with the width of the function $(\nabla_\perp^{-2} K_{d,m})(\rho', y, 0)$. We can then approximate Eq. (9) in the following way:

$$V_m\left(\sqrt{\rho^2 + y^2}\right) \cong \frac{-E(\nabla_\perp^{-2} K_{d,m})(\rho, y, 0)}{\pi d} \frac{1}{\rho} \frac{\partial}{\partial \rho} \int_\rho^{w_m} \frac{\rho' d\rho'}{(\rho'^2 - \rho^2)^{1/2}} \cong \frac{E(\nabla_\perp^{-2} K_{d,m})(\rho, y, 0)}{d\, w_m}, \tag{11}$$

where $w_m = \pi \rho_m$. It is straightforward to verify that, in the case of the Gaussian potential $V_m(\mathbf{r}) = 2Ec_m(2\pi)^{-3/2} \sigma_m^{-3} \exp[-|\mathbf{r}|^2/(2\sigma_m^2)]$, the choice $w_m = (2\pi)^{1/2} \sigma_m$ makes Eq. (11) coincide with the exact expression in Eq. (10). In general, any radial distribution $V_m(\mathbf{r})$ can be approximated by a sum of Gaussians with suitable coefficients, leading to an equation similar to Eq. (11), but with a linear combination of Gaussian terms with different coefficients $w_{mn}$ on the right-hand side [28,29]. More elaborate schemes may be unnecessary in the present context since, as we shall show below, an approximation given by Eq. (11) with a single Gaussian term is often sufficient for the purpose of reconstruction of positions and types of atoms in a molecule. Moreover, the exact value of the parameter $w_m$ is also unimportant for that purpose. Note also that because the function $(\nabla_\perp^{-2} K_{d,m})(\mathbf{r}_\perp, 0)$ is circularly symmetric in the transverse plane, the corresponding term on the right-hand side of Eq. (11) can be written as $(\nabla_\perp^{-2} K_{d,m})(\sqrt{\rho^2 + y^2}, 0, 0)$.

Adding up the atomic potentials obtained according to Eq. (11) for all atoms in the molecule, we arrive at the following expression which allows one to reconstruct the electrostatic potential from just two projections obtained by illuminating the molecule from two opposite directions at a fixed orientation:

$$V(\mathbf{r}) = \sum_{m=1}^{M} V_m(|\mathbf{r}-\mathbf{r}^{(m)}|) \cong \frac{E}{d} \sum_{m=1}^{M} w_m^{-1}(\nabla_{\perp}^{-2} K_{d,m})(|\mathbf{r}-\mathbf{r}^{(m)}|,0,0). \tag{12}$$

In practice, the effect of possible "shading" of atoms in the molecule by others along the illumination direction, as well as the signal-to-noise ratio (SNR) and sampling type considerations, imply that at least a few angular orientations are usually needed for an accurate and reliable reconstruction of a molecule. In this case, the reconstruction can be averaged over the available orientations. The 3D reconstruction, according to the algorithms outlined above, does not have to be limited to the case where the molecule is rotated around a fixed axis. In CT, it is well known that a unique 3D reconstruction of an object can be obtained from a set of projections collected for a set of different 3D orientations of the object, provided that this set satisfies the Orlov-Tuy conditions [16]. A practically important case of such type of a scan, which is also applicable to the situation considered in this paper, is represented by a set of random orientations of the molecule, where the corresponding illumination direction vectors can be considered approximately uniformly distributed on the unit sphere in 3D. The corresponding reconstruction formula can be written by analogy with Eq. (7), with the angular integration range extended to the whole unit sphere in 3D. We can write Eq. (7) for different values of the second rotational angle, $\varphi$, around the new $x_\theta$ axis (that corresponds to the position of the original $x$ axis after the initial rotation by angle $\theta$ around the $y$ axis) and average such equations over $\varphi$. This converts Eq. (7) into

$$V(\mathbf{r}) = \frac{E}{2\pi d} \nabla^{-2} \int_0^{2\pi} \int_{-\pi}^{\pi} \int_{-\infty}^{\infty} \int_{-\infty}^{\infty} \int_{-\infty}^{\infty} \exp[-i2\pi(\mathbf{r}_{\theta,\varphi}\mathbf{q}+dq_z)](\mathbf{F}_3 K_{\theta,\varphi})(\mathbf{q})|q_x|\,d\mathbf{q}\,d\theta\,d\varphi, \tag{13}$$

where an arbitrary illumination direction in 3D is represented as a result of a rotation around the $y$ axis by an angle $\theta \in [0,\pi)$ followed by a rotation around the $x_\theta$ axis by an angle $\varphi \in [0,2\pi)$:

$$\begin{cases} x_{\theta,\varphi} = x_\theta = x\sin\theta + z\cos\theta \\ y_{\theta,\varphi} = y_\theta \cos\varphi + z_\theta \sin\varphi = y\cos\varphi - x\cos\theta\sin\varphi + z\sin\theta\sin\varphi \\ z_{\theta,\varphi} = -y_\theta \sin\varphi + z_\theta \cos\varphi = -y\sin\varphi - x\cos\theta\cos\varphi + z\sin\theta\cos\varphi, \end{cases}$$

and $\mathbf{r}_{\theta,\varphi} \equiv (x_{\theta,\varphi}, y_{\theta,\varphi}, z_{\theta,\varphi})$.

A simplified approximate form of Eq. (13) can be obtained by averaging Eq. (12) over all possible illumination directions:

$$V(\mathbf{r}) \cong \frac{E}{4\pi wd} \nabla^{-2} \int_0^{2\pi} \int_{-\pi}^{\pi} K_{\theta,\varphi}(\mathbf{r}_{\theta,\varphi,\perp}, z_{\theta,\varphi} + d) \sin\varphi \, d\theta \, d\varphi, \qquad (14)$$

where $w$ is equal to the mean value of $w_m$, assuming that the widths of the circularly symmetric functions $\nabla_\perp^{-2} K_{d,m}$ are not too different from each other (which appears a reasonable assumption, for example, for molecules mainly containing C, O and N atoms). Obviously, a more accurate, although slightly more complicated, approximation of this type can be obtained by keeping different values of $w = w_m$ for different types of atoms. However, the error resulting from the use of a single parameter $w$ corresponds to a simple multiplicative correction factor $w/w_m$ for the reconstructed atomic potential $V_m(\mathbf{r})$. This error leads to over-estimation of the height of the reconstructed potential for heavier atoms, which may be unimportant when one is primarily interested only in the reconstruction of positions and types of atoms in the molecule.

Finally, in addition to arbitrary illumination directions, which are defined by the two angles $\theta$ and $\varphi$, we should allow for the possibility of the molecule to be rotated around the illumination axis $z_{\theta,\varphi}$ as well. It is straightforward to extend Eq. (14) to this case by adding another integration over the rotation angle $\psi \in [0, 2\pi)$ around the axis $z_{\theta,\varphi}$. In practice, however, attempting phase retrieval at each angle $\psi$ separately and following that with a separate backpropagation along the axis $z_{\theta,\varphi}$, is not optimal with respect to either the achievable SNR or the computational efficiency. Instead, it is more efficient to pre-process all available images corresponding to a particular illumination direction $z_{\theta,\varphi}$, by numerically rotating these images towards some fixed angle $\psi$, such as $\psi = 0$. After that, one can use all the defocused images available at a fixed illumination direction for the phase retrieval at this direction. Subsequently, the algorithm of Eq. (14) can be applied to the backpropagated contrast functions $K_{\theta,\varphi}(\mathbf{r}_{\theta,\varphi,\perp}, z_{\theta,\varphi}) \equiv K_{\theta,\varphi,\psi=0}(\mathbf{r}_{\theta,\varphi,\psi=0,\perp}, z_{\theta,\varphi})$, that is using only one 3D contrast function for each illumination direction $z_{\theta,\varphi}$.

Equation (14) yields a very straightforward and robust practical algorithm for the reconstruction of the electrostatic potential in a small sparse structure (such as a small biological molecule) from high-resolution EM images obtained at different orientations of the structure and different defocus distances. We call the algorithm described by Eq. (14) Differential Holographic Tomography (DHT). To summarize the above considerations, according to this algorithm, for each illumination direction, we reconstruct the complex wave amplitude from the collected defocused image intensities, and then numerically Fresnel back-propagate the complex amplitudes into the reconstruction volume. This allows us to evaluate the 3D distribution of the contrast function $K_{\theta,\varphi}(\mathbf{r}_{\theta,\varphi,\perp}, z_{\theta,\varphi} + d) = 1 - I(\mathbf{r}_{\theta,\varphi,\perp}, z_{\theta,\varphi} + d)/I_{in}$ on a sufficiently dense numerical grid within a virtual reconstruction volume containing the molecule, where the grid step is determined by the transverse spatial resolution of the experimental data. In accordance with Eq. (14), we then average these backpropagated 3D distributions of the contrast function over all available illumination directions, take the inverse 3D Laplacian and multiply the result by the constant factor $E/(4\pi wd)$. We will demonstrate in the next section that this approach is accurate and robust in the sense that it allows one to obtain high-quality reconstructions from a small number of defocused images of biological molecules.

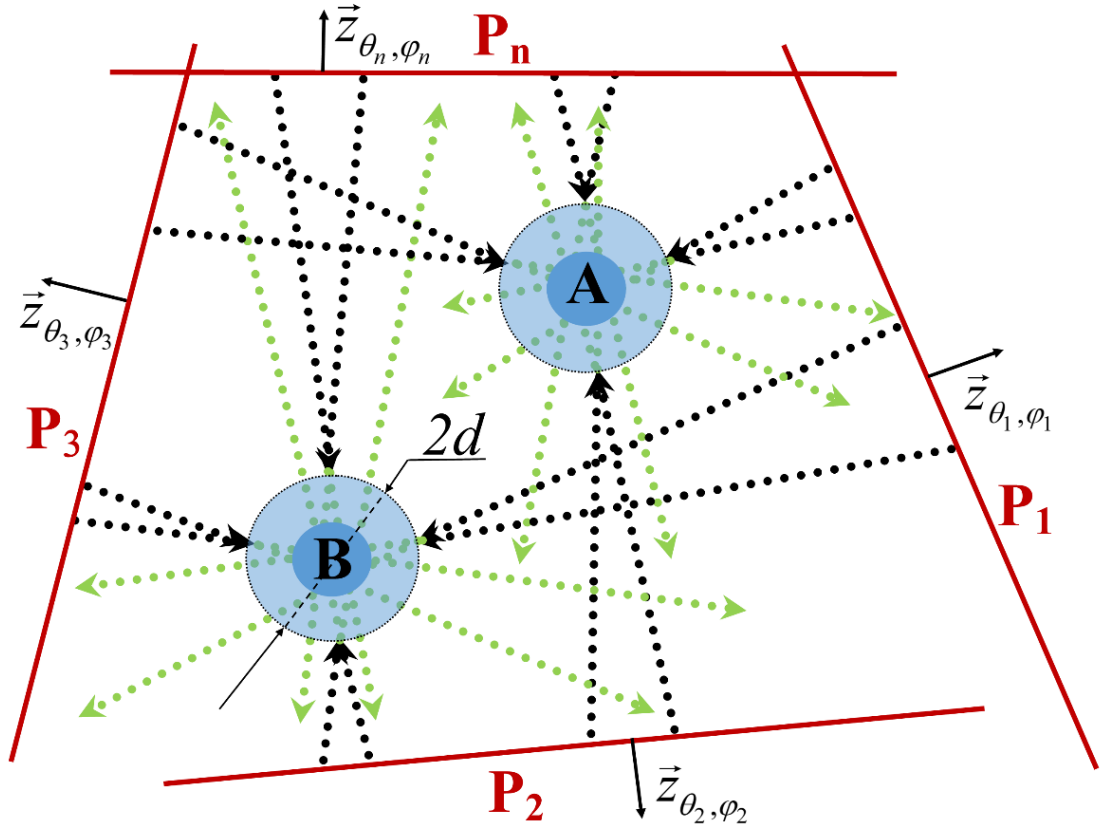

Fig. 2. Schematic representation of the DHT reconstruction algorithm.

A schematic representation of the proposed DHT algorithm is shown in Fig. 2. Multiple random illumination orientations and defocus distances correspond to planes **P**₁, **P**₂, ... and their corresponding normals $\vec{z}_{\theta_n,\varphi_n}$ (i.e. illumination directions). Only two atoms, labelled **A** and **B**, are shown. The black lines indicate the backpropagation of the reconstructed complex amplitude from planes **P**$_n$ into the reconstruction volume containing the imaged molecule. The introduced "local defocus" parameter *d* forces the backpropagated waves to converge before or after the actual *z*-location of each atom, depending on the sign of the parameter *d*. As a consequence, the introduced defocus leads to incomplete mutual cancellation of the backpropagated waves at the locations of individual atoms (which, as previously mentioned, represent pure phase objects and, therefore, do not produce intensity contrast at exact in-focus positions). Such "locally defocused" backpropagation contrast, when integrated over all available illumination directions, creates a 3D TIE-type phase contrast around each atom, which is indicated in Fig. 2 by semi-transparent blue circles. This differential phase contrast is converted into the reconstructed distribution of the electrostatic potential (shown by solid blue circles) after the application of the "phase retrieval" operator $[E/(4\pi wd)]\nabla^{-2}$ at the last step of the DHT algorithm (see eq.(14)).

## 3. Numerical simulations

In this section we describe numerical simulations performed to test the potential of practical reconstruction using the DHT method of sparse atomic structures from TEM defocus image series collected at multiple rotational orientations of the structure. The code used for the calculation of defocused images in these simulations was based on the freely available TEMSIM C++ source code developed by E.J. Kirkland [30]. The TEMSIM programs allow one to simulate defocused TEM images using the multislice method [31]. We have compared the output of our code with another well-known software package for TEM simulations, μSTEM [32], and made sure that the results of the two programs agree quantitatively with each other. Our code is also freely available at Github [33].

### *3.1. Example 1 - aspartate molecule*

The first example uses a very small biological molecule, aspartate, which contains only 16 atoms: $C_4 H_7 N O_4$ [34]. A representation of the molecule in the initial orientation used in our simulations below is shown in Fig. 3. We centred the aspartate molecule within a $10\times10\times10$ Å$^3$ cube $Q_{10}$ (where the subscript index "10" corresponds to the side-length of the cube in Å) which was located in the positive octant of the Cartesian coordinates (*x*, *y*, *z*), with one corner at the point (0, 0, 0) and all sides parallel to the coordinate axes.

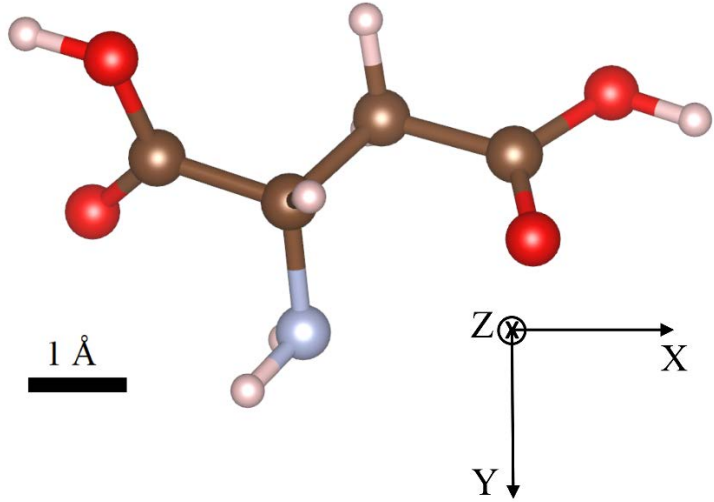

Fig. 3. Aspartate molecule as displayed by the Vesta software [35]. Oxygen atoms are red, the nitrogen atom is light blue, carbons are brown and hydrogens are light pink and small. The orientation of the axes is the same as in our simulations, with the *z* axis pointing away from the viewer.

The simulation cube $Q_{10}$ was assumed to be illuminated by a plane monochromatic incident electron plane wave with an energy of 200 keV and uniform intensity equal to one, propagating along the *z* axis. The TEMSIM code was applied for the calculation of propagation of the electron wave from the "incident" plane $(x_{\theta,\varphi}, y_{\theta,\varphi}, z_{\theta,\varphi} = 0)$ to the "exit" plane $(x_{\theta,\varphi}, y_{\theta,\varphi}, z_{\theta,\varphi} = 10\text{Å})$ at 36 different illumination directions $(\theta, \varphi)$. The unit vectors corresponding to these illumination directions were uniformly randomly distributed on the unit sphere in 3D. At each illumination direction, the calculated transmitted complex amplitude in the exit plane $z_{\theta,\varphi} = 10$ Å was rotated to three different angles $\psi_j$, $j = 1, 2, 3$, selected from a uniform random distribution of values between $\psi_{min} = 0º$ and $\psi_{max} = 360º$ (this is equivalent to rotating the molecule around the illumination axis). At each angle $\psi_j$, the rotated complex amplitude distribution was numerically propagated along the *z* coordinate (by computing the relevant Fresnel diffraction integrals) to a defocus distance $z_{\theta,\varphi} = D_j(\theta, \varphi)$ that was selected from a uniform random distribution of values contained in the interval between $D_{min} = 5$ Å and $D_{max} = 15$ Å. This procedure resulted in $36 \times 3 = 108$ defocused images corresponding to pseudo-random combinations of the four parameters, $(\theta, \varphi, \psi, D)$, selected as described above. An example of one such defocused image is shown in Fig. 4(a).

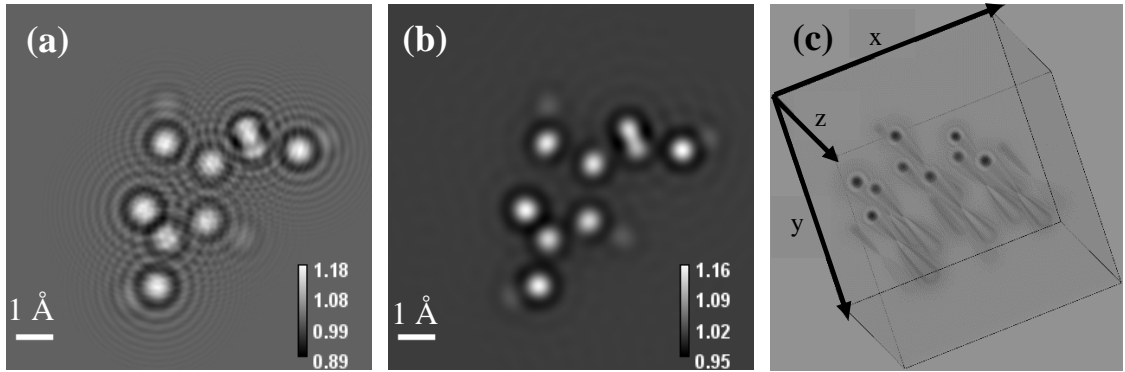

Fig. 4. Defocused images of an aspartate molecule obtained under the conditions of plane electron wave illumination (see main text for details). (a) "Ideal" image obtained at $(\theta, \varphi, \psi, D) = (26.01°, -8.52°, 129.69°, 6.74\text{Å})$; (b) image obtained at the same parameters $(\theta, \varphi, \psi, D)$, but with objective aperture of 45 mrad and thermal vibrations with 0.1 Å root-mean-square (RMS) displacement; (c) 3D rendering of a defocus series inside the cube $Q_{10}$ in the "ideal" imaging conditions at illumination angle $(\theta, \varphi) = (0,0)$. The calibration bars in (a) and (b) denote the ratio of the defocused intensity to the illuminating plane-wave intensity.

Table 1. Results of the DHT reconstruction of 3D positions of non-hydrogen atoms in an aspartate molecule from 108 defocused images calculated with 45 mrad objective aperture and 0.1 Å RMS displacement of atomic thermal vibrations. All atomic coordinates are in ångström. The "Distance" column contains the Euclidean differences (in Å) between the original and reconstructed positions of the atoms.

|  | *Atom type* | *x* | *y* | *z* | *Distance* |
|---|---|---|---|---|---|
| *Original 1* | O | 2.33 | 3.41 | 4.31 |  |
| *Reconst.1* | O | 2.32 | 3.42 | 4.37 | 0.03 |
| *Original 2* | O | 7.70 | 3.84 | 5.51 |  |
| *Reconst.2* | O | 7.69 | 3.85 | 5.55 | 0.04 |
| *Original 3* | N | 4.39 | 6.36 | 5.03 |  |
| *Reconst.3* | N | 4.38 | 6.35 | 5.06 | 0.02 |
| *Original 4* | O | 2.17 | 5.09 | 5.75 |  |
| *Reconst.4* | O | 2.17 | 5.09 | 5.78 | 0.03 |
| *Original 5* | C | 5.24 | 4.09 | 5.38 |  |
| *Reconst.5* | C | 5.24 | 4.09 | 5.42 | 0.04 |
| *Original 6* | C | 4.23 | 4.96 | 4.62 |  |
| *Reconst.6* | C | 4.24 | 4.95 | 4.68 | 0.06 |
| *Original 7* | O | 6.81 | 5.35 | 4.16 |  |
| *Reconst.7* | O | 6.82 | 5.37 | 4.18 | 0.03 |
| *Original 8* | C | 6.64 | 4.48 | 4.97 |  |
| *Reconst.8* | C | 6.63 | 4.47 | 5.01 | 0.05 |
| *Original 9* | C | 2.84 | 4.49 | 4.94 |  |
| *Reconst.9* | C | 2.84 | 4.50 | 4.98 | 0.05 |

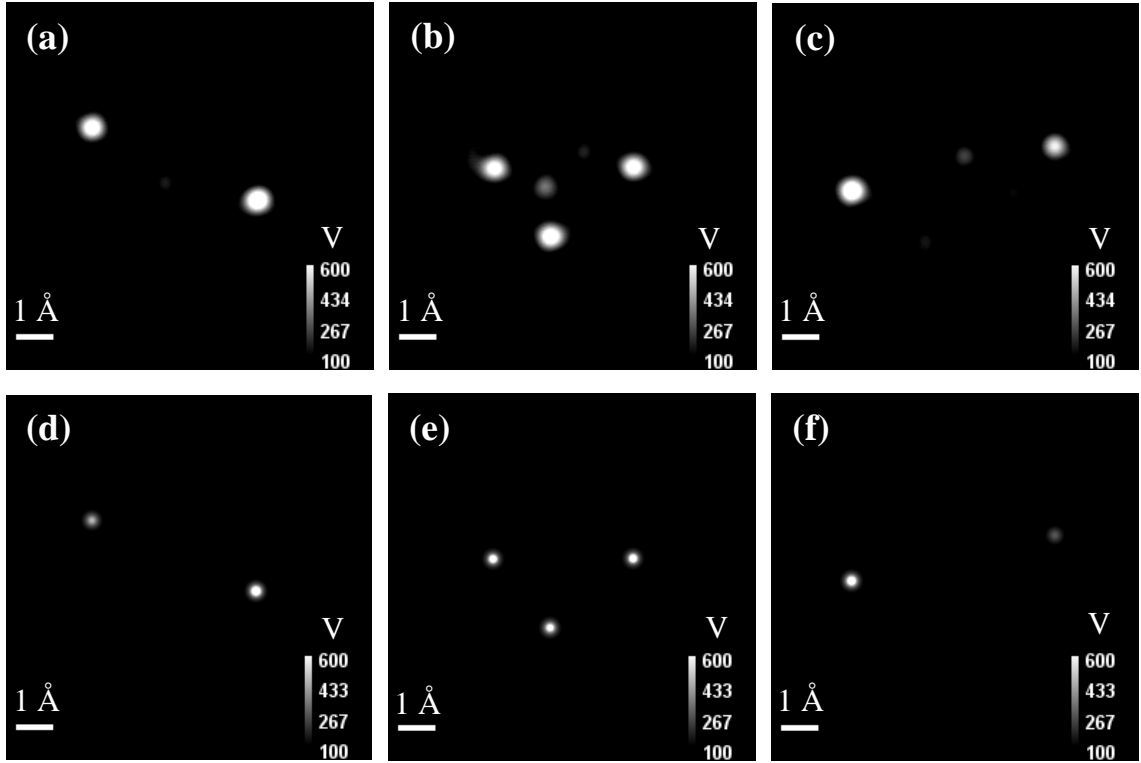

Fig. 5. Sample planar ($x, y$) cross-sections through the 3D distribution of the electrostatic potential $V(x, y, z)$ (in volt) inside the volume $Q_{10}$ containing the aspartate molecule. (a)-(c) - reconstructions from 108 simulated defocused images with objective aperture of 45 mrad and thermal vibrations with 0.1 Å RMS displacement as described in the main text (a corresponding sample defocus image is shown in Fig.4(b)); (d)-(f) - sections through the 3D spatial voltage distribution directly calculated using the modeling tool from [30]. The cross-sections correspond to the following positions: (a,d) $z = 4.22$ Å (see entries no.1 and 7 in Table 1), (b,e) $z = 5$ Å (entries no.3, 8 and 9 in Table 1), and (c,f) $z = 5.78$ Å (entries no.2 and 4 in Table 1).

Table 1 demonstrates the results of numerical reconstruction of positions and types of atoms in an aspartate molecule, using the DHT method, Eq. (14), with $d = 10$ Å, from 108 defocused images simulated at random orientations and defocus distances as described above. One can see that the positions of all O, N and C atoms have been reconstructed here with high (sub-ångström) accuracy, and all atom types have been identified correctly. The reconstruction accuracy was even better (especially in the $z$ direction) when smaller values of the local defocus parameter $d$ (e.g. $d = 1$ Å) were applied in the DHT reconstruction. However, we wanted to show here the results for a larger value of $d$, which is more suitable for defocused images affected by multiple scattering and other factors inconsistent with our simple independent atoms model based on the incoherent first Born approximation (see the results of simulations with larger molecules below).

Figure 5 presents 2D cross-sections of the reconstructed 3D distribution of the electrostatic potential $V(x, y, z)$ in different planes inside the volume $Q_{10}$ containing the aspartate molecule, together with the corresponding cross-sections of the 3D spatial distribution of the electrostatic potential directly calculated using the program "vatom" from [30] (see also section 5.2 in [31]). The reconstructed distribution was obtained using the DHT method from 108 simulated defocused images, as described above. The histograms of the reconstructed distributions shown in Fig. 5 were truncated from above and below for presentation purposes, that is to display the positions of the atoms more clearly. The reconstructed atomic positions given in Table 1 have been located as the "centres of masses" of the detected "3D objects" by the "3D Object Counter" tool (used in its default setting) in Fiji software [36]. The types of the atoms have been determined according to the values of the reconstructed potential at the corresponding positions.

### *3.2. Example 2 - lasso peptide molecule*

In the second example, we reconstruct the location of atoms in the lasso peptide (LP) molecule, $S_2O_{24}N_{23}C_{95}H_{125}$, whose structure is available in the Protein Data Bank under the code 3NJW [37]. This molecule contains 144 non-hydrogen atoms in total. We converted the description of the structure of this molecule from the standard PDB file to the XYZ file format used by TEMSIM software [30]. As part of the conversion procedure, the molecule was centered within the cube $Q_{30}$, similarly to the case with the aspartate molecule in the previous section, but with the cube $Q_{30}$ having a side length of 30 Å, as required to accommodate the LP molecule. The sampling of the latter cube consisted of $512 \times 512 \times 512$ pixels. The TEMSIM code was applied for the calculation of propagation of the electron wave from the "incident" plane $(x_{\theta,\varphi}, y_{\theta,\varphi}, z_{\theta,\varphi} = 0)$ to the "exit" plane $(x_{\theta,\varphi}, y_{\theta,\varphi}, z_{\theta,\varphi} = 30\text{Å})$ at 36 different illumination directions $(\theta, \varphi)$, with the corresponding unit vectors uniformly randomly distributed on the unit sphere in 3D. For each illumination direction, the calculated transmitted complex amplitude in the exit plane $z_{\theta,\varphi} = 30$ Å was rotated by three different angles $\psi_j$, $j = 1, 2, 3$, selected from a uniform random distribution of values between $\psi_{min} = 0°$ and $\psi_{max} = 360°$. At each angle $\psi_j$, the rotated complex amplitude distribution was numerically propagated along the $z$ coordinate to a defocus distance $z_{\theta,\varphi} = D_j(\theta, \varphi)$ that was selected from a uniform random distribution of values between $D_{min} = 5$ Å and $D_{max} = 15$ Å. This procedure resulted in 108 defocused images corresponding to exactly the same pseudo-random combinations of the four parameters, $(\theta, \varphi, \psi, D)$, as in the case of the aspartate molecule in Section 3.1. A sample defocused image is shown in Fig. 6(a). Simulations were carried out for the ideal conditions (no thermal motion and infinite objective aperture) first and then repeated with the same "realistic" conditions as used in the simulations for the aspartate molecule described in Section 3.1 (thermal motion with 0.1 Å RMS displacement and objective aperture of 45 mrad). A corresponding example image is shown in Fig. 6(b).

In both the "ideal" and the "realistic" cases, when applied to these defocused images of the lasso peptide molecule, the DHT reconstruction has performed very well (see Fig. 7). This was despite the fact that the undersampling factor with respect to angular orientations was even larger here (approaching ~ ×20) compared to the previous simulations with the aspartate molecule. In both cases, the "3D Object Counter" tool from the Fiji software [36] was able to detect all 144 non-hydrogen atoms in the reconstructed 3D TIFF image stack. The average Euclidean distance between the original positions of the atoms in the PDB file and the atom positions recovered using the 3D Object Counter tool from the DHT-reconstructed 3D distribution of the electrostatic potential was 0.067 Å in the case of "ideal" simulated defocus image data and 0.066 Å in the case of data with objective aperture of 45 mrad and thermal vibrations with 0.1 Å RMS displacement. It appears that the slightly higher accuracy was achieved in the latter case due to the smoother shape of the reconstructed electrostatic potential around individual atoms, which led to more accurate determination of the position of the centres of masses. One could assume that this achieved accuracy was superficially high because the simulations were performed with 512 numerical grid points over 30 Å distance, that is the spatial resolution could be formally regarded as being close to 0.059 Å. In reality, however, the spatial resolution in the simulated defocused images was limited by the chosen objective aperture of 45 mrad, which corresponded to the spatial resolution of approximately $2.5 \times 10^{-12}$ m / $45 \times 10^{-3} \cong 0.56$ Å for the electrons with the wavelength $\lambda = 2.5 \times 10^{-12}$ m, as used in this study.

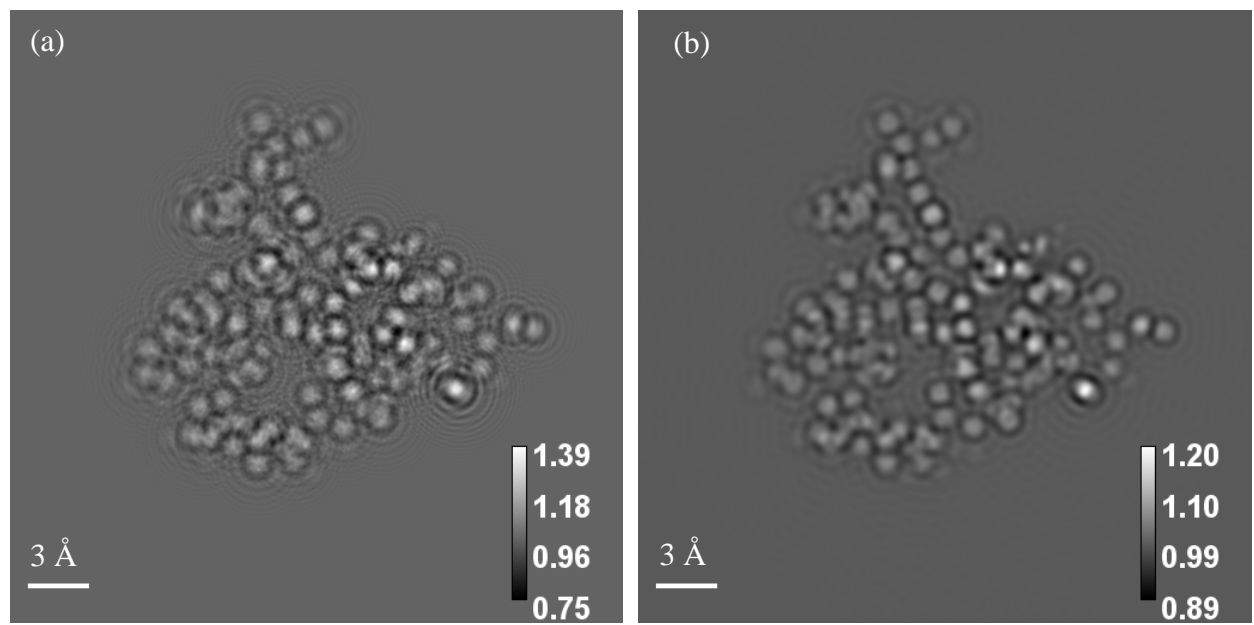

Fig. 6. Defocused images of lasso peptide 3NJW molecular structure obtained under the conditions of plane monochromatic electron wave illumination (see main text for details). (a) "Ideal" image obtained at $(\theta, \varphi, \psi, D) = (26.01°, -8.52°, 129.69°, 6.74\text{Å})$; (b) defocused image obtained at the same parameters $(\theta, \varphi, \psi, D)$, but with objective aperture of 45 mrad and thermal vibrations with 0.1 Å RMS displacement. The calibration bars denote the ratio of the defocused intensity to the illuminating plane-wave intensity.

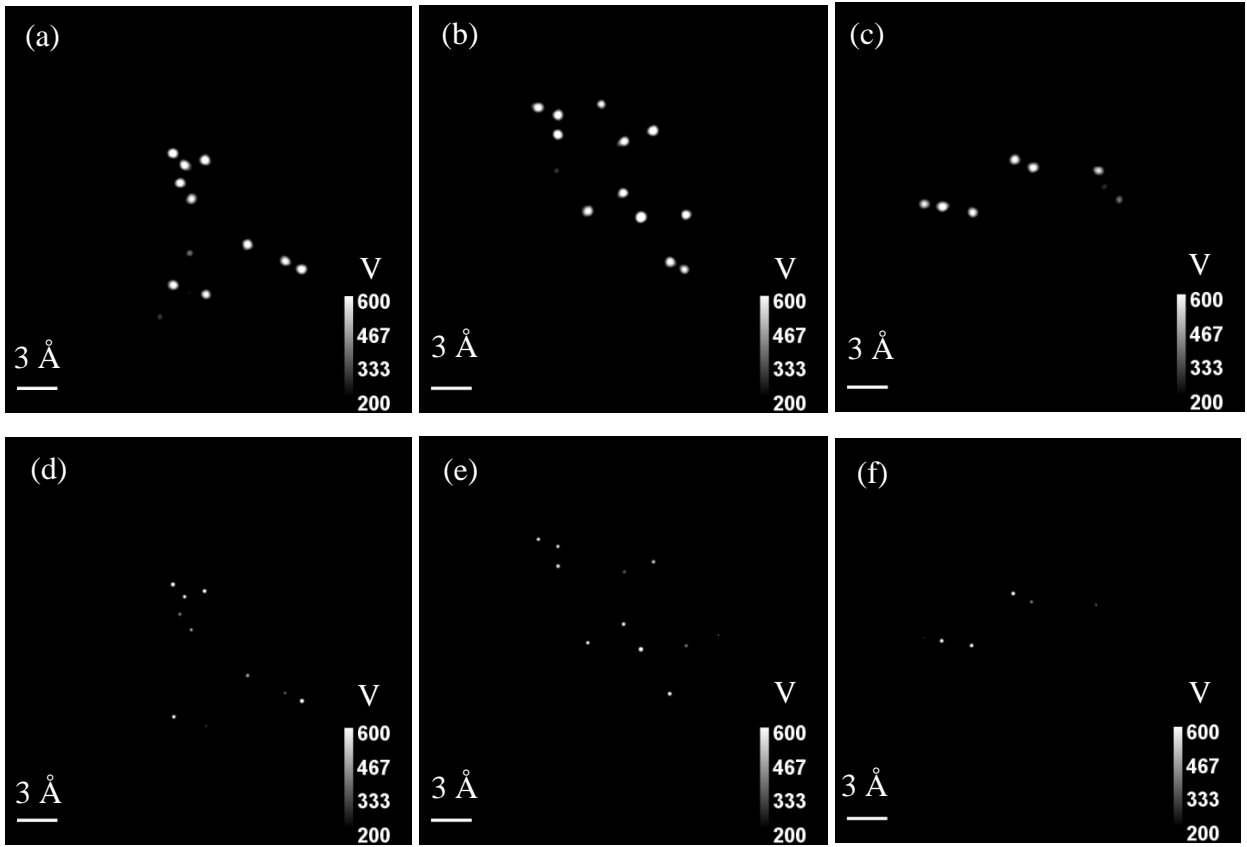

Fig. 7. Sample planar $(x, y)$ cross-sections through the 3D distribution of the electrostatic potential $V(x, y, z)$ (in volt) inside the volume $Q_{30}$ containing the lasso peptide molecule. (a)-(c) - reconstructions from 108 simulated defocused images with objective aperture of 45 mrad and thermal vibrations with 0.1 Å RMS displacement, as described in the main text (a corresponding sample defocus image is shown in Fig.6(b)); (d)-(f) - sections through the 3D spatial voltage distribution directly calculated using the modeling tool from [30]. The planes correspond to the following positions: (a, d) $z = 10$ Å, (b, e) $z = 15$ Å, and (c, f) $z = 20$ Å.

Sample cross-sections through the distributions of the reconstructed electrostatic potential $V(x, y, z)$ inside the volume $Q_{30}$ containing the lasso peptide molecule are shown in Fig. 7, together with the corresponding cross-sections of the 3D spatial distribution of the electrostatic potential directly calculated using the program "vatom" from [30]. The DHT-reconstructed potential distributions around individual atoms were lower and broader than the "ideal" theoretical ones, but, as the results of the localization of atomic positions presented above show, the local centres of masses of these broadened reconstructed distributions correlate extremely well with the exact "ideal" atomic positions.

### 3.3. Example 3 - lysozyme molecule

In some of the numerical simulations presented in our recent paper [7] we used a lysozyme molecule as an example. This molecule was the result of the default unit cell construction in Vesta software [35] from the primitive cell in the PDB file imported from the Protein Data Bank for the molecule 3J6K [38]. The lysozyme molecule (unit cell) had a volume of 214,756.23 Å$^3$ and contained 15,894 non-hydrogen atoms: $C_{9168}$ $O_{3400}$ $N_{3196}$ $S_{130}$. The PDB file did not contain the positions of hydrogen atoms and they were not used in the simulations. For these simulations, we centred the molecule within a 150×150×150 Å$^3$ cube $Q_{150}$ located in the positive octant of the Cartesian coordinates $(x, y, z)$, with one corner at the point $(0, 0, 0)$ and all sides parallel to the coordinate axes. The virtual cube $Q_{150}$ containing the lysozyme molecule was assumed to be illuminated by a plane monochromatic incident electron plane wave propagating along the $z$ axis, with an energy of 200 keV and uniform intensity equal to one.

In the "forward" stage of these simulations, the modified TEMSIM code [30] was applied for calculation of propagation of the electron wave from the "incident" plane $(x_{\theta,\varphi}, y_{\theta,\varphi}, z_{\theta,\varphi} = 0)$ to the "exit" plane $(x_{\theta,\varphi}, y_{\theta,\varphi}, z_{\theta,\varphi} = 150\text{Å})$ at different illumination directions $(\theta, \varphi)$ through the cube containing the lysozyme molecule. Each output image had a size of 150 ×150 Å$^2$ and consisted of 1024 ×1024 uniform square pixels. The objective aperture parameter was set to 45 mrad as before (corresponding to the spatial resolution of 0.56 Å), and the slice thickness in the multislice simulations was 0.1375 Å. A sample of the resultant intensity distribution in the exit plane $(x_{0,0}, y_{0,0}, 0)$ is shown in Fig. 8(a). Figures 8(b-d) show the planar $(x, y)$ cross-section, at the same longitudinal position $z = 61.23$ Å, through the 3D distribution of the electrostatic potential $V(x, y, z)$ inside the volume $Q_{150}$ containing the lysozyme molecule, reconstructed using three different methods. The pixel values in these cross-sections have been thresholded for presentation purposes from below, at $V(x, y, z) = 400$ V, and from above, at $V(x, y, z) = 600$ V, in order to remove the "background noise" and to show the potential around individual atoms as circles with discernible dimensions (rather than small dots), respectively. Figures 8(b) and (c) show the results obtained using the conventional FBP CT algorithm and the first-order DT algorithm from [7] respectively. The two reconstructions have been obtained from 180 simulated defocused images with a regular step of two degrees along the $\theta$ angle (with $\varphi = 0$), i.e. with the parameter values $(\theta, \varphi, \psi, D) = (\theta_j, 0, 0, D)$ with $\theta_j = 2^o \times j$, $j = 0, 1, ...179$, and with $D = 0$ in the case of Fig. 8(b) and $D = -15$ Å in the case of Fig. 8(c). As one can see, Fig. 8(b) contains noticeable artefacts (see e.g. the features inside the selected areas bounded by dotted circles). These artefacts originate from the Fresnel diffraction fringes and other features of the defocused images that are inconsistent with the projection approximation which serves as a basis of conventional CT (see the relevant detailed discussion in [7]). The accuracy of the DT reconstruction in Fig. 8(c) is much higher than in Fig. 8(b). The general structure of the reconstructed distribution of the electrostatic potential in Fig. 8(c) is correct and does not contain any major artefacts, except for moderate blurring of the reconstructed potential around some atomic positions (see the features inside the selected areas bounded by dashed circles), which complicates automatic localization of individual atoms. Finally, Fig. 8(d) presents the results of the reconstruction of the same transverse slice through the lysozyme molecule as shown in Figs. 8(b) and (c), but using the DHT algorithm in accordance with Eq. (14) given above. For this last result, we simulated 108 defocused images corresponding to the same pseudo-random combinations of the four parameters, $(\theta, \varphi, \psi, D)$, as in Sections 3.1 and 3.2 above, except that the defocus distances $z_{\theta,\varphi} = D_j(\theta, \varphi)$ here were uniformly randomly distributed between $D_{min}$ = -15 Å and $D_{max}$ = -5 Å. We reduced the defocus distances here, compared to Sections 3.1 and 3.2, because the "bounding box" $Q_{150}$ was significantly larger compared to the bounding boxes used in the two previous examples, and, as a consequence, the transmitted amplitude in the exit plane $z_{\theta,\varphi} = 150$ Å already displayed a significant amount of diffraction. As one can see from Fig. 8(d), the quality of the CT reconstruction using the DHT method was quite high, despite the relatively low number of orientations and defocus distances used in this example. This result confirms the statement from Section 2 above about the favorable scaling of the DHT method with respect to the number of view angles, in comparison with the usual Nyquist sampling requirements of conventional CT [16]. As explained in Section 2, the relaxation of the angular sampling requirements is due to the longitudinal localization of the back-propagated contrast function in DT around the individual atoms. After truncation of the DHT-reconstructed 3D distribution of the electrostatic potential from below, i.e. after replacing all pixel values lower than $V(x, y, z) = 400$ V by zeros, which was done in order to separate the electrostatic potential around individual atoms from the background "noise", we applied the Fiji's 3D Object Counter tool [36], in its default setting, for determination of the locations of the centres of masses of the detected "3D objects". The (*x*, *y*, *z*) positions of atoms reconstructed this way were then compared with the (*x*, *y*, *z*) positions of atoms in the original XYZ file of the lysozyme molecule used for the forward simulations. We searched for the nearest neighbor of each original atom position among the reconstructed atomic positions with respect to the Euclidean distance. The location of 15,556 out of the total of 15,894 atoms in the lysozyme molecule (i.e. 98 % of all atoms) have been correctly identified as a result, with the average distance between the original and the reconstructed atomic positions being 0.16 Å.

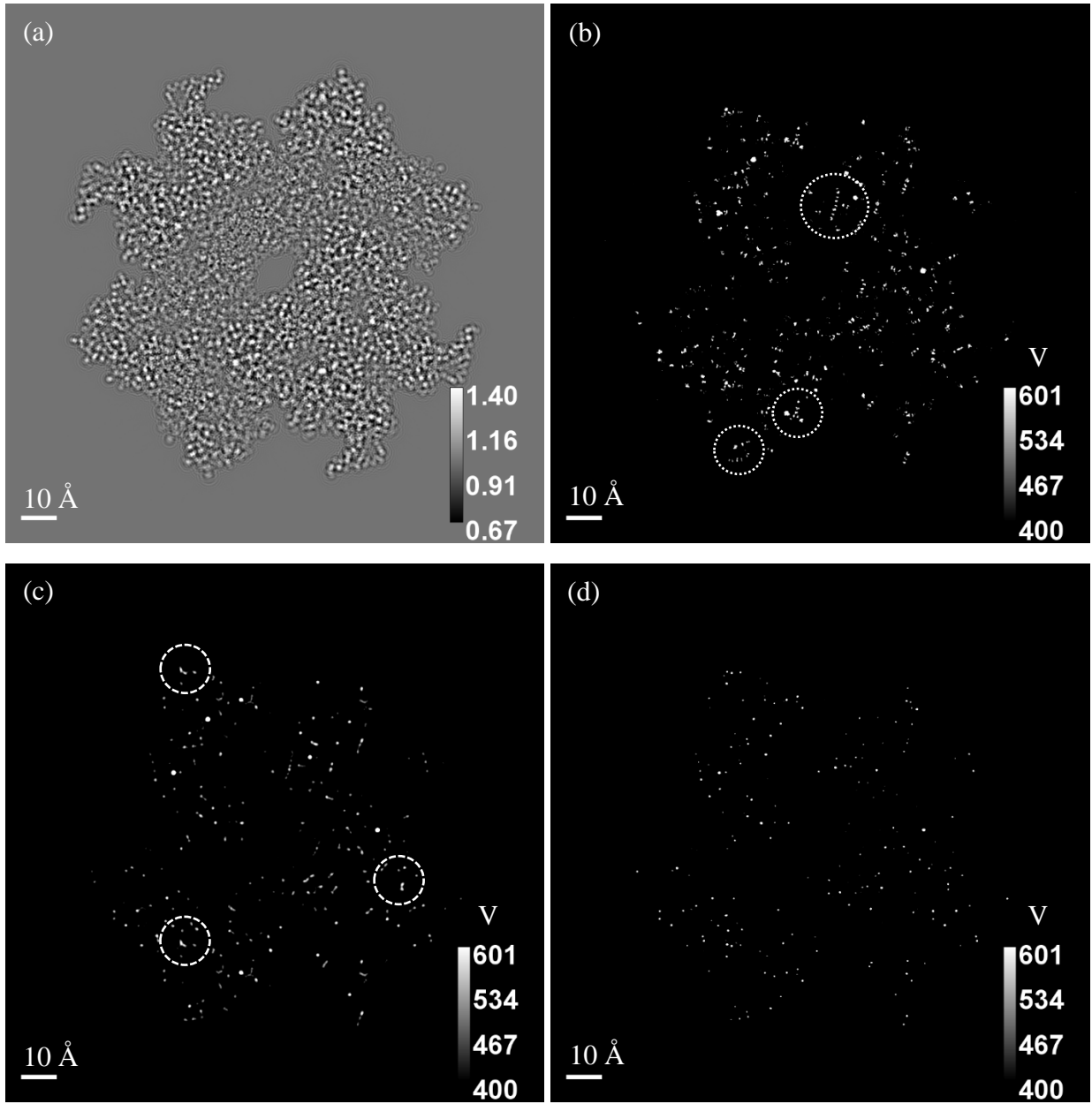

Fig. 8. Simulated defocused image and tomographic reconstructions of a lysozyme molecule (Protein Data Bank structure 3J6K [38]) using a monochromatic plane electron wave with energy $E = 200$ keV and the objective aperture of 45 mrad (see more details in the main text).
(a) Defocused image obtained at $(\theta, \varphi, \psi, D) = (0,0,0,0)$. (b-d) Thresholded planar $(x, y)$ cross-section at $z = 61.23$ Å through the 3D distribution of the electrostatic potential $V(x, y, z)$ (in volt) inside the volume $Q_{150}$ containing the lysozyme molecule, reconstructed by three different methods. Dotted and dashed circles identify regions with obvious reconstruction artefacts.
(b) Conventional FBP CT reconstruction from 180 regularly spaced projections.
(c) Reconstruction using the first-order DT algorithm from [7] from 180 regularly spaced projections. (d) DHT reconstruction from 108 defocused images simulated at pseudo-randomly distributed rotational orientations and defocus distances (see main text for details).

## 5. Conclusions

In the first part of this work [7], we argued that in TEM imaging of small biological molecules or other "sparsely localized" weakly scattering structures, multiple scattering tends to have only a moderate effect and therefore can be safely ignored in the reconstruction procedures, without introducing large errors into the results. Importantly, the in-molecule free-space propagation (Fresnel diffraction) cannot be ignored because of the shallow depth of focus under typical TEM imaging conditions. As DT represents precisely the technique which takes into account the free-space propagation, but not the multiple scattering between different atoms, it appears to be a very good match for this case, as has been argued by other authors previously [26,39]. However, acquisition of images on a regular angular grid over the full $2\pi$ range of rotation within a fixed plane, as required in DT, can often be challenging in high-resolution EM. As one way to alleviate this problem, in the present paper we have developed the technique of Differential Holographic Tomography (DHT), which can reconstruct the 3D distribution of the electrostatic potential in a molecule from defocused images collected at arbitrary orientations of the molecule and arbitrary defocus distances. This method fully exploits the information about the 3D atom locations which is available in TEM defocus series. Such information is inherently present in typical cryo-EM data. The DHT method beneficially utilizes this information resulting in improved localization of atomic positions compared to conventional CT, and in the significantly reduced requirements for the number of different view angles (rotational positions of the sample) required for unambiguous reconstruction of the sample.

In the current implementation of the DHT method, the IWFR algorithm was used for phase retrieval, i.e. for the reconstruction of the complex amplitude, at each available illumination direction. This procedure requires two or more images to be collected at different defocus distances at each illumination direction. However, as we mentioned earlier, different methods for phase retrieval, e.g. the method presented in reference [14], may allow the reconstruction of the complex wave amplitude from a single defocused image. Moreover, it is possible in principle to perform the numerical backpropagation of each defocused image into the reconstruction volume, as required in the DHT method, without any explicit phase retrieval procedure. This can be achieved, for example, by using the same first Born approximation for a pure phase object as was used in Section 2 above for the description of the forward propagation of the scattered electron wave. The issue that needs to be carefully dealt with in such an approach is the well-known problem of division by zeros of the CTF. Various approaches have been suggested in the past in order to deal with this problem [23,40-42]. Note that, in the context of the DHT method, the IWFR phase retrieval can be also viewed as a regularization technique that stabilizes the subsequent numerical backpropagation operation with respect to zeros of the CTF.

We have demonstrated, by means of numerical experiments using multislice calculations for the simulation of TEM defocus series, that the DHT method can be expected to perform well under realistic imaging conditions. Apart from cryo-EM, the DHT method can potentially be applied to other forms of EM tomography of small sparse structures, such as e.g. high-resolution imaging of nano-particles [43].

## Acknowledgements

The authors wish to thank Dr. Andrew Martin for sharing his software code that was used in the course of this research.

## Appendix A. TEM contrast function for sparse structures with Gaussian "atomic" potentials

Consider the model case where the sample has a sparse and sharply localized structure, i.e. the distribution of the electrostatic potential in the sample can be represented in the form

$$V(\mathbf{r}) = 2E \sum_{m=1}^{M} c_m \delta(\mathbf{r}_\perp - \mathbf{r}_\perp^{(m)}) \delta(z - z^{(m)}), \tag{A1}$$

where $\mathbf{r}^{(m)} \equiv (\mathbf{r}_\perp^{(m)}, z^{(m)})$ are the positions of individual "atoms" and $2Ec_m$ are the "integral values" of the potential at each such position (the coefficients $c_m$ have dimensionality m$^3$). Taking a 2D Fourier transform of Eq. (A1) and substituting the result into Eq. (1), we obtain
$(\mathbf{F}_2 I)(\mathbf{q}_\perp, z) / I_{in} = \delta(\mathbf{q}_\perp) - (4\pi/\lambda) \sum_{m=1}^{M} c_m \sin[\pi\lambda(z^{(m)} - z)q_\perp^2] \exp(-i2\pi\mathbf{q}_\perp \mathbf{r}_\perp^{(m)})$. The 2D inverse Fourier transform of this expression is equal to

$$I(\mathbf{r}_\perp, z) / I_{in} = 1 - (4\pi/\lambda) \sum_{m=1}^{M} c_m \int d\mathbf{q}_\perp \exp[i2\pi\mathbf{q}_\perp(\mathbf{r}_\perp - \mathbf{r}_\perp^{(m)})] \sin[\pi\lambda(z^{(m)} - z)q_\perp^2]$$

$$= 1 + i(2\pi/\lambda) \sum_{m=1}^{M} c_m \int d\mathbf{q}_\perp \exp[i2\pi\mathbf{q}_\perp(\mathbf{r}_\perp - \mathbf{r}_\perp^{(m)})] \{\exp[i\pi\lambda(z^{(m)} - z)q_\perp^2] - \exp[-i\pi\lambda(z^{(m)} - z)q_\perp^2]\}.$$

The latter expression corresponds to 2D Fresnel diffraction integrals and can be easily evaluated (see Appendix A4 in [44]), with the resulting contrast function

$$K(\mathbf{r}) = \frac{4\pi}{\lambda^2} \sum_{m=1}^{M} c_m \frac{\cos\left[\dfrac{\pi |\mathbf{r}_\perp - \mathbf{r}_\perp^{(m)}|^2}{\lambda(z - z^{(m)})}\right]}{z - z^{(m)}}. \tag{A2}$$

Note that the expression on the right-hand side of Eq. (A2) is dimensionless. The general structure of this equation already correctly reflects some of the key features of typical defocus series of sparse atomic structures, including the contrast sign reversal at points $z = z^{(m)}$. However, the contrast function in Eq. (A2) has singularities at these contrast reversal points, which is a direct consequence of the singularity of the delta functions in Eq. (A1). In order to make the model for $V(\mathbf{r})$ more realistic and get rid of the singularities, we now convolve each term on the right-hand side of Eq. (A1) with a Gaussian $G^{(m)}(\mathbf{r}) \equiv g^{(m)}(x)g^{(m)}(y)g^{(m)}(z)$, where $g^{(m)}(x) \equiv (2\pi\sigma_m^2)^{-1/2} \exp[-x^2/(2\sigma_m^2)]$, etc., so that

$$V(\mathbf{r}) = 2E\sum_{m=1}^{M} c_m G_\perp^{(m)}(\mathbf{r}_\perp - \mathbf{r}_\perp^{(m)}) g^{(m)}(z - z^{(m)}), \qquad (A3)$$

with $G_\perp^{(m)}(\mathbf{r}_\perp) \equiv g^{(m)}(x)g^{(m)}(y)$. Because of the linearity of Eq. (1) with respect to $V(\mathbf{r})$, we can now simply convolve each term on the right-hand side of Eq. (A2) with the corresponding Gaussian in order to obtain the relevant expression for the contrast function using the model of the potential from Eq. (A3). With respect to integration over $z$ in these convolutions, we follow the approach used in the multislice method [27] and replace the $z$-convolutions of the propagator terms $[z - z^{(m)}]^{-1} \cos[(\pi/\lambda)|\mathbf{r}_\perp - \mathbf{r}_\perp^{(m)}|^2/(z - z^{(m)})]$ with $g^{(m)}(z)$ by the products of the value of the propagator term at point $z$ and the $z$-projected value of the potential, which is equal to 1, since $\int g^{(m)}(z)dz = 1$. The convolutions with respect to $x$ and $y$ can be calculated explicitly, since they correspond to Fresnel propagation by distances $\pm(z - z^{(m)})$ of the two-dimensional Gaussians:

$$K(\mathbf{r}) = \frac{1}{\lambda^2}\sum_{m=1}^{M} \frac{c_m}{\sigma_m^2(z-z^{(m)})}\int d\mathbf{r}_\perp' \exp\left[\frac{-|\mathbf{r}_\perp'|^2}{2\sigma_m^2}\right]\exp\left[\frac{i\pi|\mathbf{r}_\perp - \mathbf{r}_\perp^{(m)} + \mathbf{r}_\perp'|^2}{\lambda(z-z^{(m)})}\right] + c.c., \qquad (A4)$$

where "$c.c$" means the complex conjugate of the preceding expression. Note that the different treatment of the convolutions with respect to $z$ and $\mathbf{r}_\perp$ are justified here in view of the paraxial approximation adopted in this problem (corresponding to a situation dominated by small angle scattering).

Using the known expression for paraxial propagation of Gaussian beams [45], we obtain directly from Eq. (A4):

$$K(\mathbf{r}) = \frac{4}{\lambda}\sum_{m=1}^{M}\frac{c_m}{w_m^2(z)}\exp\left[\frac{-|\mathbf{r}_\perp - \mathbf{r}_\perp^{(m)}|^2}{w_m^2(z)}\right]\sin\left[\frac{\Delta\tilde{z}_m|\mathbf{r}_\perp - \mathbf{r}_\perp^{(m)}|^2}{w_m^2(z)} - \arctan(\Delta\tilde{z}_m)\right], \qquad (A5)$$

where $\Delta \tilde{z}_m \equiv \lambda (z - z^{(m)}) / (2\pi \sigma_m^2)$ and $w_m^2(z) \equiv 2\sigma_m^2 [1 + (\Delta \tilde{z}_m)^2]$. The main properties of the contrast function in Eq. (A5) are as follows.

(i) The sine factor in each term under the sum in Eq. (A5) describes the oscillatory profile of the propagated intensity. This factor also makes the contrast function change its sign along the $z$ coordinate at the point $z = z^{(m)}$, as a consequence of the definition of $\Delta \tilde{z}_m = \lambda (z - z^{(m)}) / (2\pi \sigma_m^2)$ and the fact that both sine and arctan functions are odd. Related to this change of sign is also the fact that the contrast produced by a given atom with index $m$ is equal to zero in the plane $z = z^{(m)}$ (which is a consequence of the weak phase approximation used in the present model). If the atomic structure is sparse, the contribution of atoms with indices $m' \neq m$ to the contrast function at point $(\mathbf{r}_\perp^{(m)}, z^{(m)})$ is weak (see below), and hence the point of disappearance of the contrast can be easily located in the through-focus series and the spatial positions $\mathbf{r}^{(m)} = (\mathbf{r}_\perp^{(m)}, z^{(m)})$ of individual atoms can be located by this means.

The type of the atom located at position $\mathbf{r}^{(m)}$ is determined by the value of $c_m$, in the model of Eq. (A3). This value can be found from the magnitude of the change of the contrast function between the positions $(\mathbf{r}_\perp^{(m)}, z^{(m)} - \varepsilon)$ and $(\mathbf{r}_\perp^{(m)}, z^{(m)} + \varepsilon)$ for a small defocus $\varepsilon$. The arctan term in Eq. (5) correspond to the Gouy phase [46].

(ii) The exponential factor in each term under the sum in Eq. (A5) determines the strong transverse localization of the contrast function around the axis $\mathbf{r}_\perp = \mathbf{r}_\perp^{(m)}$, parallel to $z$, making each term in the sum behave as a Gaussian beam with waist radius $w_0^2 \equiv w_m^2(0) = 2\sigma_m^2$, achieved in the plane $z = z^{(m)}$. The longitudinal localization (i.e. the localization along the $z$ coordinate) of the contrast function in Eq. (A5) is determined primarily by the factor
$w_m^{-2}(z) \equiv 0.5 \sigma_m^{-2} [1 + (\Delta \tilde{z}_m)^2]^{-1} = 2\pi^2 \sigma_m^2 / [(2\pi \sigma_m^2)^2 + \lambda^2 (z - z^{(m)})^2]$. Therefore, the diffraction contrast decreases along $z$ generally in proportion to the square of the distance from the plane in which the atom is located. In this sense, the longitudinal localization of the "traces" of atoms in the defocus series is weaker than their transverse localization and the structure needs to be more sparse in the $z$ direction, compared to $(x, y)$, in order to avoid the "screening" of the defocused images of some atoms by other atoms located on the same or adjacent lines $z$ = constant.

(iii) Note that the parameters $c_m$ are not independent of the parameters $\sigma_m$, as both of them are determined by the type of a given atom. If the atomic structure in Eq. (A3) is sparse, the $4M$ independent parameters, e.g. $(c_m, \mathbf{r}^{(m)})$, $m = 1, ..., M$, can be found from 2D images collected at defocus distances around $z = z^{(m)}$. At such defocus positions, generally only the atom located at $(\mathbf{r}_\perp^{(m)}, z^{(m)})$ will contribute non-negligibly to the contrast function around this point, since the contribution from all other terms in Eq. (A5) will be strongly attenuated by the small factors $w_m^{-2}(z)\exp[-w_m^{-2}(z)|\mathbf{r}_\perp - \mathbf{r}_\perp^{(m)}|^2]$, if the sparsity condition is satisfied and the abovementioned "screening" can be avoided.

The corresponding picture in reciprocal space can be obtained by substituting the 3D Fourier transform of Eq. (A3) into Eq. (1), expressing the sine function under the integral sign in Eq. (1) via a difference of two complex exponents and then taking the 1D Fourier transform with respect to $z$:

$$(\mathbf{F}_3 K)(\mathbf{q}) = (i2\pi/\lambda)\sum_{m=1}^{M} c_m \exp\{-2\pi^2\sigma_m^2 q_\perp^2[1+(\lambda/2)^2 q_\perp^2]\}\exp(-i2\pi\mathbf{r}_\perp^{(m)}\mathbf{q}_\perp)$$
$$\times\{\exp(-i\pi\lambda z^{(m)}q_\perp^2)\delta[q_z - (\lambda/2)q_\perp^2] - \exp(i\pi\lambda z^{(m)}q_\perp^2)\delta[q_z + (\lambda/2)q_\perp^2]\}.$$
(A6)

Note that Eq. (A6) represents a 3D Fourier transform of intensity registered in the Fresnel region, rather than the intensity of a complex wave in the reciprocal space, which can be obtained, for example, in the far-field diffraction regime. This equation shows that, at a fixed angular orientation of the molecule, the non-zero contrast is limited to the paraboloids $q_z = \pm(\lambda/2)q_\perp^2$. For all atoms, the contrast exponentially decreases as a function of reciprocal coordinates, according to the term $\exp\{-2\pi^2\sigma_m^2 q_\perp^2[1+(\lambda/2)^2 q_\perp^2]\}$. However, since the Gaussians in Eq. (A5) are supposed to be narrow, with widths of approximately 1 Å, the latter exponential terms are going to be broad in the reciprocal space, with widths of the order of 1 Å$^{-1}$. Therefore, the diffraction "signal" from different atoms in the model of Eq. (A3) will be sufficiently strongly represented on the support of the delta functions in Eq. (A6). The corresponding phases depend on the projection of the atom location vector $(\mathbf{r}_\perp^{(m)}, z^{(m)})$ onto the reciprocal vectors $\mathbf{q}$, according to the term $\exp\{-i2\pi[\mathbf{r}_\perp^{(m)}\mathbf{q}_\perp \mp z^{(m)}(\lambda/2)q_\perp^2]\}$. The information about the atoms' types and 3D locations can be extracted from this type of data, in principle.

Note that Eq. (A5) can be obtained by the inverse 3D Fourier transform of Eq. (A6), provided that we ignore the terms $(\lambda/2)^2 q_\perp^2$ in the first exponent. If we assume, for example, that the desired spatial resolution is close to 1 Å and the electron wavelength is around 0.025 Å, then we get $(\lambda/2)^2(q_{x,\max}^2 + q_{y,\max}^2) < 2\times 10^{-4}$. Therefore, in such cases, $(\lambda/2)^2 q_\perp^2 \ll 1$ and hence the approximation required for the derivation of Eq. (A5) from Eq. (A6) is well justified.